\documentclass[a4paper,12pt]{article}
\usepackage{amsmath}
\usepackage{amssymb}
\usepackage{listings}
\usepackage{comment}

\usepackage{subfigure}
\def\be#1{\begin{equation}\label{#1}}
\def\ee{\end{equation}}
\def\eq#1{(\ref{#1})}

\newcommand {\Oe}       [1]     {$$}

\newcommand {\ba}       [2]     {\be{#1} \begin{array}{#2}}
\newcommand {\ea}               {\end{array} \ee}
\newcommand {\Oa}       [2]     {$$ \begin{array}{#2}}
\def\oa{\end{array} $$}

\let\DS=\displaystyle

\def\({\left(}
\def\){\right)}

\usepackage{ifpdf}

\ifpdf
  \usepackage[pdftex]{graphicx}
  \DeclareGraphicsExtensions{.png,.jpg,.tif}
\else
  \usepackage{graphicx}
  \DeclareGraphicsExtensions{.bmp,.pcx}
\fi

\newlength{\pix}
\newlength{\grhh}\newlength{\grww}

\iftrue     

\else

\fi

\def\Xi{\Tens{\xi}{}}
\def\Eta{\Tens{K}{}}

\def\Gamma{\Tens{\gamma}{}}

\def\dev{{\rm dev}}

\def\rr{\Vect{r}}

\def\Tpi{\Tens{\Pi}}

\def\EE{\Tens{E}}

\def\Nu{\Tens{q}}

\def\LL{\Tens{L}}

\def\Teps{\Tens{H}}

\def\D{\Tens{\cal{D}}}

\def\vv{\Vect{v}}

\def\va{\Vect{a}_{\alpha}}
\def\eal{\Vect{e}_{\alpha}}
\def\R3{{}^3\!\Tens{R}}
\def\S4{{}^4\!\Tens{S}}

\let\<=\langle \let\>=\rangle
\def\av#1{\Bigl\langle{#1}\Bigr\rangle}

\iffalse
    \def\Tens#1{{\underline{\underline{#1\hspace{-0.5mm}}}\hspace{0.5mm}}}
    \def\Vect#1{{\underline{#1\hspace{-0.5mm}}\hspace{0.5mm}}}
\else
    \newcommand{\Vect}[1]{\boldsymbol{\mathrm{#1}}{}}
    \newcommand{\Tens}[1]{\boldsymbol{\mathrm{#1}} {}}
\fi

\def\tr{\mathop{\rm tr}}

\def\u{\Vect u}

\def\eal{\Vect e_\al}

\def\R{\Vect R}

\def\eal{\Vect{e}_{\alpha}}

\def\paptit#1{}

\begin{document}

\title{High-frequency thermal processes in harmonic crystals}

\author{Vitaly A. Kuzkin 
               \and Anton M. Krivtsov\footnote{Peter the Great Saint Petersburg Polytechnical University,
              Polytechnicheskaya st. 29, Saint Petersburg, Russia;
              Institute for Problems in Mechanical Engineering RAS, Bolshoy pr. V.O. 61, Saint Petersburg, Russia;
              e-mails: kuzkinva@gmail.com (V.A. Kuzkin), akrivtsov@bk.ru (A.M. Krivtsov) }
}

\maketitle

\begin{abstract}
We consider two high frequency thermal processes in uniformly heated harmonic crystals relaxing towards equilibrium: (i) equilibration of kinetic and potential energies and (ii) redistribution of energy among spatial directions. Equation describing these processes with deterministic initial conditions is derived. Solution of the equation shows that characteristic time of these processes is of the order of ten periods of atomic vibrations. After that time the system practically reaches the stationary state. It is shown analytically that in harmonic crystals temperature tensor is not isotropic even in the stationary state. As an example, harmonic triangular lattice is considered. Simple formula relating the stationary value of the temperature tensor and initial conditions is derived. The function describing equilibration of kinetic and potential energies is obtained. It is shown that the difference between the energies~(Lagrangian) oscillates around zero. Amplitude of these oscillations  decays inversely proportional to time. Analytical results are in a good agreement with numerical simulations.

\end{abstract}

\section{Introduction}
Description of nonequilibrium thermal processes is a challenging problem for modern mechanics and physics of solids. The problem is particularly important due to recent advances in nanotechnologies~\cite{Goldstein 2007}. In the present paper, fast thermal processes~\cite{Krivtsov 2014 DAN, Krivtsov_Tropp,Babenkov} accompanying the transition of the system from nonequilibrium state towards thermodynamic equilibrium are considered~\cite{Evans}. Initial nonequilibrium state is caused, for example, by fempto- or attosecond laser excitation~\cite{laser} or by shock waves~\cite{Hoover Holian}. In this state kinetic and potential energies are not equal. Also kinetic temperatures corresponding to thermal motion of atoms in different spatial directions may be different~\cite{Hoover Holian}.\footnote{Here and below kinetic temperature proportional to kinetic energy of thermal motion is considered.} Computer simulations show that in harmoniñ crystals there are two thermal processes accompanying the transition to equilibrium:(i) equilibration of kinetic and potential energies~\cite{Krivtsov 2014 DAN, Krivtsov_Tropp, Allen} and (ii) redistribution of energy among spatial directions. Characteristic time of these processes is of the order of ten periods of atomic vibrations.\footnote{We define the period as~$\tau_*=2\pi\sqrt{m/C}$, where $m$ is the mass of the atom, $C$ is the bond stiffness.}

The present paper focuses on the analytical description of the above mentioned thermal  processes. The approach described in papers~\cite{Krivtsov 2014 DAN, Babenkov} is generalized for the multidimensional case. Two- and three-dimensional harmonic crystals with random initial velocities and displacements are considered. Thermal processes in the crystal are described using the correlation analysis~\cite{Krivtsov 2014 DAN, Babenkov, Rieder 1967, Krivtsov 2015 DAN, Krivtsov 2015 Arxive}. Deterministic problem for the generalized energies~\cite{Krivtsov 2014 DAN,  Krivtsov_Tropp, Babenkov, Krivtsov 2015 DAN, Krivtsov 2015 Arxive} is formulated. Solution of the deterministic problem yields the function describing equilibration of kinetic and potential energies and the relation between temperatures in different spatial directions in the stationary state.

\section{An analytical description of transient thermal processes}
We consider an infinite simple crystal lattice consisting of particles with equal masses. The nearest neighbors interact via linearized~(harmonic) forces. Particles are identified by their radius-vectors in the undeformed state. Motion of particles  is described by differential-difference equation\footnote{This equation is equivalent to the infinite system of ordinary differential equations of the second order.}:
\be{EM}
\begin{array}{l}
\DS \dot{\vv}(\rr) = \D \cdot \u(\rr), \qquad
 \D = \omega_*^2\sum_{\alpha}  \eal\eal\Delta_{\alpha}^2, \\[4mm]
 \DS \Delta_{\alpha}^2\u(\rr) = \u(\rr+\va)-2\u(\rr)+\u(\rr-\va),
\end{array}
\ee
where $\u(\rr)$, $\vv(\rr)$ are displacement and velocity of the particle with radius-vector~$\rr$; $\D$ is the tensor difference operator; $\va$ is the vector connecting two neighboring particles; $\eal = \va/|\va|$; $\omega_*=\sqrt{C/m}$; $C$ is the bond stiffness; $m$ is the particle's mass. The summation is carried out over noncollinear  bond directions~$\alpha$. In particular,~$\alpha=1,2$ for a square lattice and ~$\alpha=1,2,3$ for a triangular lattice. Initial displacements and velocities of the particles have the form:
\be{IC}
 t=0: \qquad \u(\rr) = \u_0(\rr), \quad \vv(\rr) = \vv_0(\rr),
\ee
where $\u_0$, $\vv_0$ are independent random vectors with zero expected value.

Equation~\eq{EM} with initial conditions~\eq{IC} completely determines the dynamics of the crystal. In principle, it can be solved analytically.  The resulting solution yields stochastic displacements and velocities of all particles. In contrast, description of macroscopic thermal processes usually focuses on statistical characteristics such as temperature.

In papers~\cite{Krivtsov 2014 DAN, Krivtsov_Tropp, Krivtsov 2015 DAN, Krivtsov 2015 Arxive} it is shown  that kinetic temperature is insufficient for obtaining closed system of equations. A closed system can be derived for the generalized energies. We define the generalized~(two-particle) kinetic~$\Eta$ and potential~$\Tpi$ energies for particles~$i$ and $j$ with radius-vectors~$\rr_i$ and $\rr_j$ as:
\be{}
\begin{array}{l}
\DS\Eta(\rr_i,\rr_j)\!=\!\frac{m}{2} \av{\vv_i \vv_j},
\qquad
\Tpi(\rr_i,\rr_j)\!=\! -\frac{m}{4}\(\D\!\cdot\!\av{\u_i\u_j} + \av{\u_i\u_j}\!\cdot\!\D\).
\end{array}
\ee
Here~$\vv_i = \vv(\rr_i)$; brackets~$\av{}$ denote the expected value of a random variable. For~$i=j$, the traces of tensors~$\Eta$, $\Tpi$ correspond to conventional kinetic and potential energies per particle. We also define the {\it tensor temperature}~$\Tens{T}$~\cite{Hoover Holian} as
\be{Tens T}
\frac{k_B}{2} \Tens{T}(\rr_i) = \frac{m}{2}\av{\vv_i \vv_i}  = \Eta|_{i=j}, \qquad T = \frac{1}{d}\tr\Tens{T},
\ee
where $k_B$ is Boltzmann's constant, $d$ is the space dimensionality, $T$ is the conventional kinetic temperature. The generalized total energy~$\Teps$ and
the generalized Lagrangian~$\LL$ are defined as
\be{K P}
\Teps = \Eta + \Tpi, \qquad \LL = \Eta - \Tpi.
\ee

Below processes in uniformly heated crystals are considered. In this case the generalized energies satisfy the relations~$\Eta(\rr_i,\rr_j)=\Eta(\rr_i-\rr_j)$, $\Tpi(\rr_i,\rr_j)=\Tpi(\rr_i-\rr_j)$. Argument~$\rr_i-\rr_j$ is omitted below for brevity. Note that the points defined by vectors~$\rr_i-\rr_j$ form the same lattice as vectors~$\rr_i$.

We show that the generalized total energy~$\Teps$ satisfies several conservation laws. Computing the derivative of~$\Teps$
and using equations of motion~\eq{EM} yields:
\be{dot E}
\DS \dot{\Teps} = \D \cdot \Nu - \Nu \cdot \D, \qquad \Nu\!=\!\frac{m}{4}\av{\u_i \vv_j - \vv_i\u_j}.
\ee
Multiplying equation~\eq{dot E} by $\D^n$, we obtain the conservation laws
\be{const_n}
 \D^n\cdot\cdot\Teps = {\rm const}, \qquad n=0,1,2,...
\ee
In the case~$n=0$ and $i=j$ formula~\eq{const_n} coincides with the conventional law of energy conservation. From the Cayley--Hamilton theorem it follows that the number of independent conservation laws~\eq{const_n} is equal to space dimensionality.

 We derive the dynamic equations for generalized energies. Differentiating the generalized Lagrangian~$\LL$ with respect to time and taking equations of motion~\eq{EM} into account yields:
\be{Dyn En}
\begin{array}{l}
\DS \ddddot{\LL} - 2\(\D\cdot\ddot{\LL} + \ddot{\LL}\cdot\D\) + \D^2\cdot\LL-2\D\cdot\LL\cdot\D + \LL\cdot\D^2 = 0,
\end{array}
\ee
where $\D^2=\D\cdot\D$. It can be shown that quantities~$\Eta$, $\Tpi$, $\Teps$ also satisfy equation~\eq{Dyn En}.
 Corresponding initial conditions are uniquely determined by initial displacements and velocities of the particles~\eq{IC}. Thus dynamics of the generalized energies is described by equation~\eq{Dyn En} with {\it deterministic} initial conditions.

Computer simulations show that after a short transient process the system practically reaches the {\it stationary state}.\footnote{The stationary state is defined so that the second time derivatives of the generalized energies are equal to zero.} Consider the relation between stationary values of the generalized energies and the initial conditions.  Using equation~\eq{Dyn En} for~$\Teps$ and conservation laws~\eq{const_n} yields the closed system of
equations for stationary values of the generalized energies:
\be{dev Eta}
\begin{array}{l}
\DS \tr \Teps  = \tr\Teps_0,
\quad
\D \cdot\cdot \Teps   = \D \cdot\cdot \Teps_0,
\quad
\D^2 \cdot\cdot \Teps   = \D^2\cdot\cdot \Teps_0,
\\[4mm]
\DS    \D^2\cdot\Teps-2\D\cdot\Teps\cdot\D + \Teps\cdot\D^2 = 0,
\end{array}
\ee
where $\Teps_0$ is the initial value of the generalized total energy. The relation between the generalized kinetic and potential energies follows from the identity~$\frac{m}{4}\av{\u_i \u_j}\ddot{} =  \LL$.
In the stationary state, the left hand side of this equation is equal to zero, then
\be{K=P}
 \Eta  = \Tpi  = \frac{1}{2} \Teps.
\ee
Equation~\eq{K=P} and the first of equations~\eq{dev Eta} lead to the following expression for traces of the generalized energies:~$\tr \Eta  = \tr\Tpi  = \frac{1}{2}\tr\Teps_0$.
In the particular case~$i=j$, the given expression follows from the virial theorem. Deviators of the generalized energies are determined by solution of equations~\eq{dev Eta}, \eq{K=P}.

\section{Example: harmonic triangular lattice}

The derivations presented above are  valid in two-  and three-dimensional cases. As an example, consider transition to the stationary state in the two-dimensional harmonic triangular lattice. Initial velocities of the particles are independent random vectors with zero mean; initial displacements are equal to zero. In this case initial conditions for the Lagrangian~$\LL$ have the form:
\be{IC L}
\begin{array}{l}
t=0: \qquad \LL = \Teps = \Eta_0\delta(\rr_i-\rr_j),
\quad
\Eta_0 = \frac{m}{2} \av{\vv_i\vv_i},
\quad
\dot{\LL} = 0, \\[4mm]
\DS \ddot{\LL} = 2\(\D\cdot\LL+\LL\cdot\D\), \quad  \dddot{\LL} = 0,
\end{array}
\ee
where $\delta(\rr_i-\rr_j)=0$ for~$i\neq j$;  $\delta(0) = 1$.

Consider the stationary value of the tensor temperature~\eq{Tens T}.  Equations~\eq{dev Eta} with initial conditions~\eq{IC L} are solved with respect to~$\Tens{H}$ using  the discrete Fourier transform. After that the tensor temperature is calculated using formulas~\eq{Tens T}, \eq{K=P}.
 For an infinite triangular lattice the following relation
 between the temperature tensor~\eq{Tens T} and initial conditions is obtained:
\be{Klaw}
  k_B\,\dev\Tens{T}  = \frac{1}{4}\,\dev\Eta_0.
\ee
Therefore the temperature tensor in the stationary state, in general, is not isotropic.

\begin{figure*}[htb]
\begin{center}
\includegraphics*[scale=0.5]{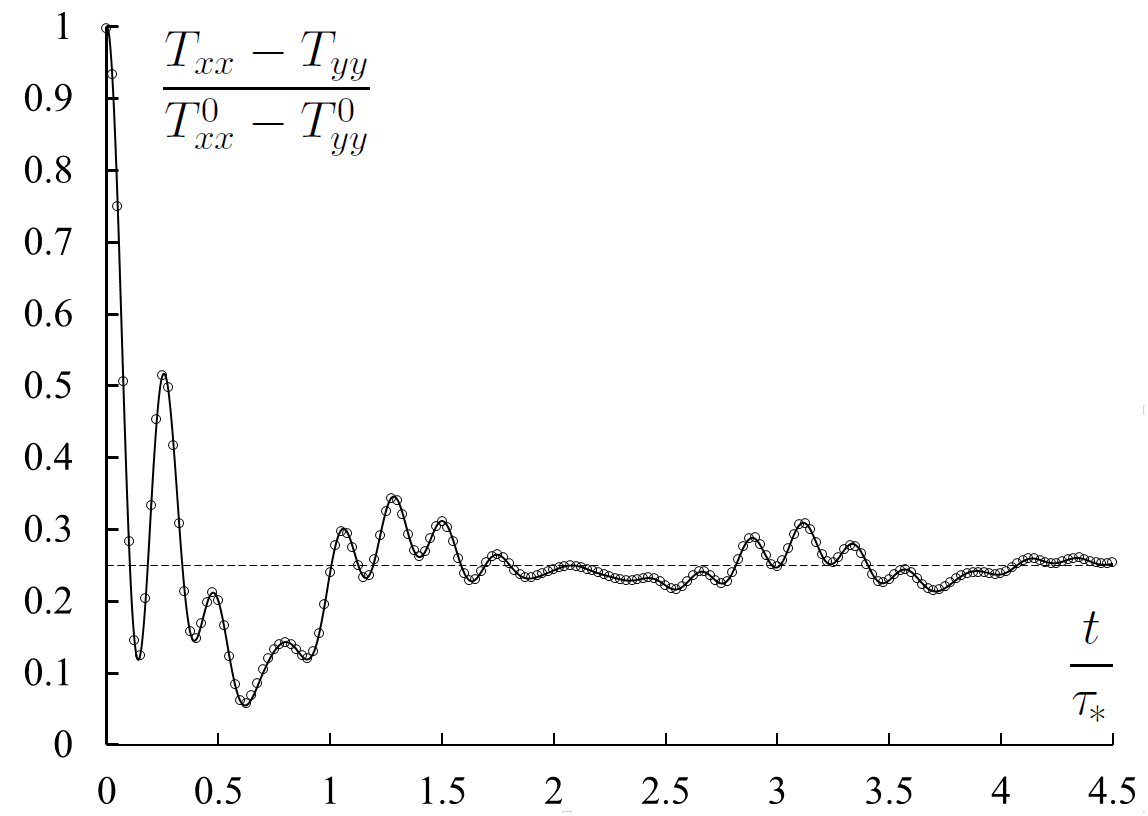}
\caption{Time-dependence of the difference of temperatures corresponding to~$x$ and $y$ directions. Circles  --- numerical solution of lattice dynamics equations \eq{EM} ($1.6\cdot 10^4$ particles; periodic boundary conditions). Solid line --- numerical solution of the differential-difference equation~\eq{Dyn En} for the generalized kinetic energy. Dashed line --- solution of stationary problem~\eq{Klaw}.}
\label{Eta_xx-Eta_yy}
\end{center}
\end{figure*}

We compare the results given by formula~\eq{Klaw} with numerical solution of lattice dynamics equations~\eq{EM}. Verlet integration scheme with the time step~$10^{-3} \tau_*$ is used. Here and below~$\tau_* =2\pi/\omega_*$ . Initial displacements of particles are equal to zero. Initial velocities have random magnitude and directed along one of the lattice directions~($x$ axis).  Time dependence of the difference between temperatures in~$x$ and $y$ directions is shown in figure~\ref{Eta_xx-Eta_yy}. Numerical solution of equation~\eq{Dyn En} for~$\Eta$ is also given for comparison.

%

It is seen that the system practically reaches the stationary state after several periods~$\tau_*$. In the stationary state, the relation~\eq{Klaw} is satisfied. Transition to the stationary state is exactly described by equation~\eq{Dyn En}.

\begin{figure*}[htb]
\begin{center}
\includegraphics*[scale=0.5]{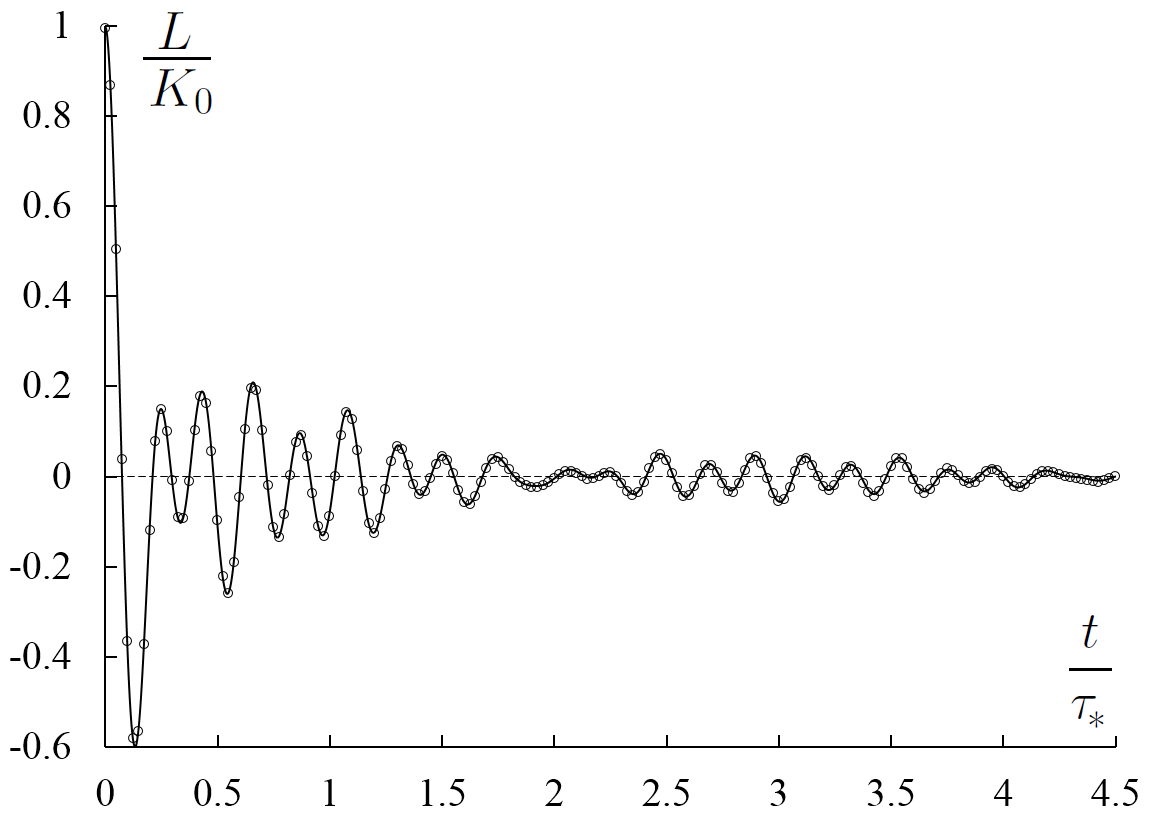}
\caption{Oscillations of the Lagrangian in the crystal with random initial velocities. Solid line --- analytical solution~\eq{en_osc}; circles --- numerical solution of lattice dynamics equations~\eq{EM}.}
\label{Txx_Tyy_tensor_vs_MD}
\end{center}
\end{figure*}

Consider equilibration of kinetic and potential energies in the triangular lattice. The process is described by equation~\eq{Dyn En}. Initial particle velocities are independent random vectors, uniformly distributed among spatial directions. In this case~$\Eta_0 = \frac{K_0}{2}\EE$, where $\EE$ is the unit tensor.
The following assumption is used for solution of equation~\eq{Dyn En} with initial conditions~\eq{IC L}\footnote{This assumption significantly simplifies equation~\eq{Dyn En}. Comparison with numerical solution shows that equation~\eq{en_sym} correctly describes behavior of Lagrangian~$L=\tr\LL|_{i=j}$~(see Fig.~\ref{Txx_Tyy_tensor_vs_MD}). Therefore the assumption is acceptable.}
~$\D\cdot\LL = \LL \cdot \D$. Then equation~\eq{Dyn En} takes the form
\be{en_sym}
  \ddot{\LL} = 4\D\cdot \LL.
\ee
Note that equation~\eq{en_sym} is similar to the equation of motion~\eq{EM}. Equation~\eq{en_sym} is solved using the discrete Fourier transform. In the case of infinite crystal the solution
gives the following expression for Lagrangian~$L=\tr\LL|_{i=j}$:
\be{en_osc}
  \begin{array}{l}
\DS L(t) = \frac{K_0}{2\pi^2}\int_{0}^{\pi}\!\!\int_{0}^{\pi}\!\(\cos(2\Omega_1(s,p)t) + \cos(2\Omega_2(s,p)t)
\){\rm d} s{\rm d}p,
\\[5mm]
\DS \Omega_{k}^4 - 4\omega_*^2\Omega_{k}^2(\sin^2s + \sin^2p + \sin^2(s+p)) + 12\omega_*^4\( \sin^2s\sin^2p +  \right. \\[4mm]
\DS  \left. \sin^2(s+p)(\sin^2s + \sin^2p)\) = 0, \qquad k=1,2.
\end{array}
\ee
The second formula from~\eq{en_osc} corresponds to the dispersion relation for the triangular lattice.

Formula~\eq{en_osc} shows that the difference between kinetic and potential energies oscillates with the  amplitude inversely proportional to time. It decays by two orders of magnitude in~$10\tau_{*}$. In the one-dimensional case,  similar oscillations are described by the Bessel function of the first order, which decays inversely proportional to the square root of time~\cite{Krivtsov 2014 DAN}.

Analytical solution~\eq{en_osc} is compared with numerical solution of lattice dynamics equations~\eq{EM}~(see Fig.~\ref{Txx_Tyy_tensor_vs_MD}). It is seen that solutions practically coincide.

\section{Conclusions}
Thus in the present paper the analytical description of two nonequilibrium thermal processes, notably (i) equilibration of kinetic and potential energies and (ii) redistribution of energy among spatial directions, was proposed. Equation~\eq{Dyn En} with deterministic initial conditions describing both processes in two- and tree-dimensional cases was derived. Stationary values of the generalized energies are related with the initial conditions by equations~\eq{dev Eta}, \eq{K=P}. It was shown that in the triangular lattice the temperature tensor is not isotropic.
Its deviator in the stationary state is determined by formula~\eq{Klaw}:
\be{}
    T_{xx}-T_{yy} = \frac{1}{4}\(T_{xx}^0-T_{yy}^0\),
\ee
where $T_{xx}^0$, $T_{yy}^0$ are initial temperatures in~$x$ and $y$ directions.
Also it was shown that equilibration of kinetic and potential energies is described by function~\eq{en_osc}.

The results obtained in the present paper  can be used for description of fast thermal processes in weakly anharmonic crystals~(at low temperatures). In paper~\cite{Benettin} it is shown that small nonlinearity leads to slow energy exchange between the normal modes.  As a result, at short times considered above the effect of nonlinearity is weak.

Also the results may serve for description of anomalous heat transfer. This process in defect-free crystals, in general, is not described by the Fourier law~\cite{Krivtsov_Tropp, Rieder 1967, Krivtsov 2015 DAN, Krivtsov 2015 Arxive, Lepri, Dhar 2008, SavinGendelman}.

The authors are deeply grateful to M.B. Babenkov W.G. Hoover, and D.A. Indeytsev for useful discussions.



\end{document}